\documentclass[review]{elsarticle}

\usepackage{lineno}
\usepackage{hyperref}
\modulolinenumbers[5]
\usepackage{graphicx}
\usepackage{epstopdf, epsfig, amsmath}
\usepackage{algorithm2e}
\usepackage{physics}
\usepackage{amssymb}
\usepackage{xcolor}
\usepackage{soul}
\usepackage{pstricks}
\usepackage{arydshln}

\newcommand\qvec{\mathbf{q}}
\newcommand\qfft{\hat{\mathbf{q}}}
\newcommand\Qmatfft{\hat{\mathbf{Q}}}

\newcommand\avec{\mathbf{a}}
\newcommand\bvec{\mathbf{b}}
\newcommand\xvec{\mathbf{x}}

\newcommand\uvec{\mathbf{u}}
\newcommand\vvec{\mathbf{v}}

\newcommand\Umat{\mathbf{U}}
\newcommand\Vmat{\mathbf{V}}
\newcommand\Xmat{\mathbf{X}}

\newcommand\Kmat{\mathbf{K}}

\newcommand\Mmat{\mathbf{M}}
\newcommand\Qmat{\mathbf{Q}}
\newcommand\Smat{\mathbf{S}}
\newcommand\Wmat{\mathbf{W}}
\newcommand\Imat{\mathbf{I}}
\newcommand\Zmat{\mathbf{0}}
\newcommand\phivec{\boldsymbol{\phi}}
\newcommand\SigmaMat{\boldsymbol{\Sigma}}

\newcommand\PhiMat{\boldsymbol{\Phi}}
\newcommand\LambdaMat{\boldsymbol{\Lambda}}
\newcommand\realNum{\mathbb{R}}

\newcommand\nblk{{n_\text{blk}}}
\newcommand\nfreq{{n_\text{freq}}}

\newcommand\nvar{{n_\text{var}}}
\newcommand\novlp{{n_\text{ovlp}}}
\newcommand{\qmean}{\overline{\qvec}}
\newcommand{\ii}{\mathrm{i}}

\begin{document}

\begin{frontmatter}

\title{An efficient streaming algorithm for spectral proper orthogonal decomposition}

\author[1]{Oliver T. Schmidt\corref{mycorrespondingauthor}}
\cortext[mycorrespondingauthor]{Corresponding author}
\ead{oschmidt@ucsd.edu}
\author[2]{Aaron Towne}

\address[1]{University of California San Diego, La Jolla, CA 92093, USA }
\address[2]{University of Michigan, Ann Arbor, MI 48109, USA}

\begin{abstract}
A streaming algorithm to compute the spectral proper orthogonal decomposition (SPOD) of stationary random processes is presented. As new data becomes available, an incremental update of the truncated eigenbasis of the estimated cross-spectral density (CSD) matrix is performed. The algorithm requires access to only one temporal snapshot of the data at a time and converges orthogonal sets of SPOD modes at discrete frequencies that are optimally ranked in terms of energy. We define measures of error and convergence, and demonstrate the algorithm's performance on two datasets. The first example considers a high-fidelity numerical simulation of a turbulent jet, and the second uses optical flow data obtained from high-speed camera recordings of a stepped spillway experiment. For both cases, the most energetic SPOD modes are reliably converged. The algorithm's low memory requirement enables real-time deployment and allows for the convergence of second-order statistics from arbitrarily long streams of data.
\end{abstract}


\end{frontmatter}


\section{Introduction}\label{intro}

The ability to represent complex dynamics by a small number of dynamically important modes enables the analysis, modeling and control of high-dimensional systems. Turbulent flows are a prominent example of such systems \cite{taira2017modal, rowley2017model}. Proper orthogonal decomposition (POD), also known as principle component analysis (PCA) or Karhunen-Lo\`{e}ve decomposition, is a popular modal decomposition technique to extract coherent structures from experimental and numerical data. In its most common form \cite{sirovich1987turbulence}, POD is conducted in the time domain. It is computed from a time series of snapshots and expands the flow field into a sum of products of spatially orthogonal modes and coefficients with random time dependance. POD modes are optimally ranked in terms of their variance, or energy. These properties make POD modes well suited for low-order models based on Galerkin projection of the Navier-Stokes equations \cite{aubry1988dynamics,noack2003hierarchy}.

Besides its definitions in the temporal and spatial domains, POD can also be formulated in the frequency domain. This variant of POD called \emph{spectral proper orthogonal decomposition} (SPOD), dates back to the early work of Lumley \cite{Lumley:1970} and takes advantage of temporal homogeneity. This makes it ideally suited for statistically (wide-sense) stationary data \cite{towneschmidtcolonius_2018_jfm}. SPOD provides orthogonal modes at discrete frequencies that are optimally ranked in terms of energy, and that evolve coherently in both space and time. Perhaps predictably, this optimal space-time representation of the data comes at a cost--very long time series are necessary in order to converge the second-order space-time statistics. This data demand also becomes apparent when comparing SPOD to other modal decomposition techniques, for example the popular dynamic mode decomposition (DMD) \cite{schmid2010dmd}. For stationary data, SPOD modes correspond to optimally averaged DMD modes computed from an ensemble of stochastic realizations of a process \cite{towneschmidtcolonius_2018_jfm}, for example multiple repetitions of the same experiment. In \S \ref{examples}, we use two databases consisting of 10,000 and 19,782 snapshots, respectively. Evidently, the SPOD problem quickly becomes computationally unmanageable for data with large spatial dimensions.

In this paper, we address this issue by proposing a low-storage \emph{streaming SPOD} algorithm  that incrementally updates the SPOD as new data becomes available. Similar algorithms are often referred to as \emph{incremental}, \emph{learning}, \emph{updating}, \emph{on-the-fly} or \emph{online} algorithms in the literature. Streaming algorithms for DMD have been developed recently \cite{hemati2014dynamic,zhang2017online}, for example. The proposed streaming SPOD algorithm utilizes incremental updates of the singular value decomposition (SVD) of the cross-spectral density (CSD) matrix of the data. SVD updating has been an active research topic for almost half a century, see e.g.  \cite{businger1970updating,de1985svd}. In this work, we build on Brand's \cite{brand2006fast} incremental singular value decomposition (SVD) by specializing the method to updates of the estimated CSD matrix. Originally developed for computer vision and audio feature extraction \cite{brand2002incremental}, the algorithm has been employed for recommender systems \cite{brand2003fast}, semantics \cite{turney2010frequency}, design optimization \cite{braconnier2011towards}, and a wide spectrum of other machine learning and data mining applications.

The paper is organized as follows. We first introduce standard or \emph{batch} SPOD in \S\ref{batch} before deriving the streaming algorithm in \S\ref{onlinespod}. Measures of error and convergence are defined in \S\ref{errors}. In \S\ref{examples}, we demonstrate streaming SPOD on two datasets: a high-fidelity large eddy simulation (LES) of a turbulent jet and experimental optical flow from high-speed camera data of a stepped spillway. The effect of eigenbasis truncation is addressed in \S \ref{rankderror}. In \S\ref{discussion}, we conclude with a discussion of the algorithm's computational efficiency and utility in real-time and big data settings.

\section{Batch SPOD}\label{batch}
SPOD is the frequency-domain counterpart of standard time-domain or spatial POD. SPOD yields time-harmonic modes that represent structures that evolve coherently in both time and space \cite{towneschmidtcolonius_2018_jfm}. The method is based on an eigendecomposition of the CSD, which in turn is estimated from an ensemble of realizations of the temporal discrete Fourier transform (DFT) in practice. The CSD can be estimated using standard spectral estimation techniques such as Welch's method \cite[see e.g.][]{Solomon1991} from an ensemble of snapshots. The SPOD formalism is derived from a space-time POD problem under the assumption of wide-sense stationarity. The reader is referred to \cite{towneschmidtcolonius_2018_jfm} for the derivation of the method and an assessment of its properties. In particular, the method's relations to DMD and the resolvent operator are interesting from a modeling perspective, as they link SPOD to concepts from dynamical systems and hydrodynamics stability theory.

\begin{figure}
\centering
	\input{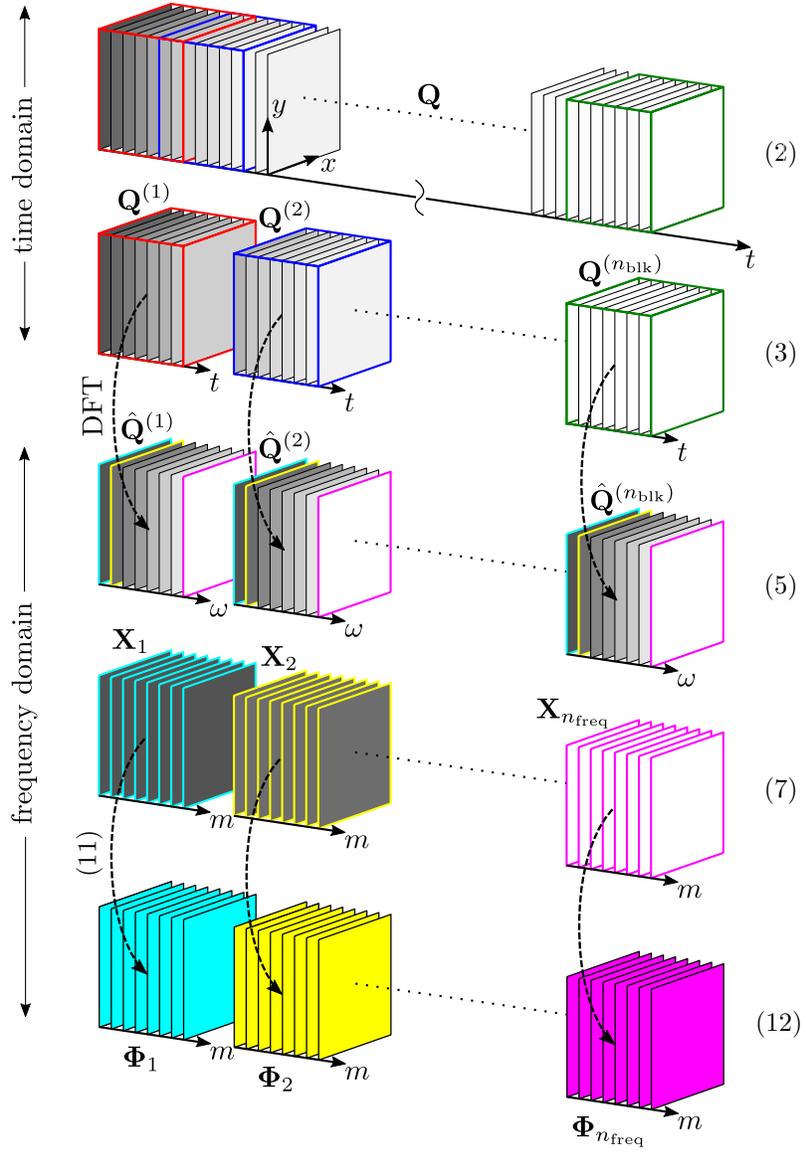}
	\caption{Illustration of the batch SPOD algorithm. Each rectangular slice represents a snapshot and the numbers in parentheses denote the equations in the text. The data is first segmented, then Fourier transformed, then reordered by frequency, and finally diagonalized into SPOD modes.}
	\label{fig:SPODsketch}
\end{figure}
Figure \ref{fig:SPODsketch} serves as a visual guide through the batch algorithm. We start with an ensemble of $n_t$ snapshots $\qvec_i=\qvec(t_i)\in\realNum^n$ of a wide-sense stationary process $\qvec(t)$ sampled at discrete times $t_1,t_2,\dots,t_{n_t}\in\realNum$. By $\qvec$ we denote the state vector. Its total length $n$ is equal to the number of grid points $n_x$ times the number of variables $\nvar$. The temporal mean corresponds to the ensemble average defined as
\begin{equation}
\qmean = \frac{1}{n_t} \sum_{i=1}^{n_t} \qvec_i  \in\realNum^{n}.
\end{equation}
We collect the mean-subtracted snapshots in a data matrix 
\begin{equation}
\Qmat=[\qvec_1-\qmean,\qvec_2-\qmean,\dots,\qvec_{n_t}-\qmean]\in\realNum^{n\times {n_t}}
\end{equation}
of rank $d\leq\min\{n,{n_t-1}\}$. With the goal of estimating the CSD, we apply Welch's method to the data by segmenting $\Qmat$ into $\nblk$ overlapping blocks 
\begin{equation}
\Qmat^{(l)}=[\qvec_1^{(l)}-\qmean,\qvec_2^{(l)}-\qmean,\dots,\qvec_\nfreq^{(l)}-\qmean]\in\realNum^{n\times \nfreq}
\end{equation}
containing $\nfreq$ snapshots each. If $\novlp$ is the number of snapshots by which the blocks overlap, then the $j$-th column of the $l$-th block $\Qmat^{(l)}$ is given as
\begin{equation}
\label{eqn:snapshot_placement}
\qvec_j^{(l)}=\qvec_{j+(l-1)(\nfreq-\novlp)}-\qmean.
\end{equation}

We assume that each block can be regarded as a statistically independent realization under the ergodicity hypothesis. The purpose of the segmentation step is to artificially increase the number of ensemble members, i.e.~Fourier realizations. This method is useful in the common scenario where a single long dataset with equally sampled snapshots is available, for example from a numerical simulation. In situations where the data presents itself in form of independent realizations from the beginning, segmenting need not be applied. This is the case, for example, if an experiment is repeated multiple times. Next, the temporal (row-wise) discrete Fourier transform
\begin{equation} \label{eqn:dft_block}
\Qmatfft^{(l)}=[\qfft_1^{(l)},\qfft_2^{(l)},\dots,\qfft_\nfreq^{(l)}]\in\realNum^{n\times \nfreq}
\end{equation}
of each block is calculated. A windowing function can be used to mitigate spectral leakage. All realizations of the Fourier transform at the $k$-th frequency are subsequently collected into a new data matrix
\begin{equation}
\Qmatfft_k=[\qfft_k^{(1)},\qfft_k^{(2)},\dots,\qfft_k^{(\nblk)}]\in\realNum^{n\times \nblk}.
\end{equation}
At this point, we introduce the weighted data matrix
\begin{equation}\label{eqn:xmat}
\Xmat_k=\frac{1}{\sqrt{\nblk}}\Wmat^{\frac{1}{2}}\Qmatfft_k = [\xvec_k^{(1)},\xvec_k^{(2)},\dots,\xvec_k^{(\nblk)}]\in\realNum^{n\times \nblk},
\end{equation}
where $\Wmat\in\realNum^{n\times n}$ is a positive-definite Hermitian matrix that accounts for quadrature and possibly other weights associated with the discretized inner product 
\begin{equation}\label{eqn:innerp}
\left<\avec,\bvec\right>_E = \avec^*\Wmat\bvec.
\end{equation}
The inner product (\ref{eqn:innerp}) induces the spatial energy norm $\|\cdot\|_E=\sqrt{\left<\cdot,\cdot\right>_E}$ by which we wish to rank the SPOD modes. The product
\begin{equation}\label{eqn:csd}
\Smat_{k}=\Xmat_k\Xmat_k^*\in\realNum^{n\times n}
\end{equation}
defines the weighted CSD matrix of the $k$-th frequency. A factor of $\frac{1}{\nblk}$ seen in other definitions of the CSD is absorbed into our definition of the weighted data matrix in equation (\ref{eqn:xmat}). 

SPOD is based on the eigenvalue decomposition
\begin{equation}
\Smat_{k}= \Umat_k \LambdaMat_k \Umat_k^*
\end{equation}
of the CSD matrix, where $\LambdaMat_k = \mathrm{diag}(\lambda_{k_1},\lambda_{k_2}\cdots,\lambda_{k_\nblk}) \in \realNum^{\nblk\times \nblk}$ is the matrix of ranked (in descending order) eigenvalues and $\Umat_k = [\uvec_{k_1},\uvec_{k_2},\dots,\uvec_{k_\nblk}] \in \realNum^{n\times \nblk}$ the corresponding matrix of eigenvectors. Equivalently, we may consider the economy SVD of the weighted data matrix
\begin{equation}\label{eqn:svdofqhat}
\Xmat_k = \Umat_k
\SigmaMat_k
\Vmat_k^*,
\end{equation}
where $\SigmaMat_k = \mathrm{diag}(\sigma_{k_1},\sigma_{k_2}\cdots,\sigma_{k_\nblk})  \in \realNum^{\nblk\times \nblk}$ is the matrix of singular values and $\Vmat_k = [\vvec_{k_1},\vvec_{k_2},\dots,\vvec_{k_\nblk}] \in \realNum^{n\times \nblk}$ the right singular vector matrix. This can be shown by rewriting the CSD in terms of the SVD of the data matrix as $\Smat_{k}= \Xmat_k\Xmat_k^* = \Umat_k \SigmaMat_k \Vmat_k^* \Vmat_k \SigmaMat_k \Umat_k^* = \Umat_k \LambdaMat_k \Umat_k^*$. Throughout this paper, we assume that all Fourier realizations of the flow are linearly independent. In the final step, the SPOD modes $\phi$ and modal energies $\sigma^2$ are found as 
\begin{eqnarray}\label{eqn:batchmodes}
\PhiMat_k=\Wmat^{-\frac{1}{2}}\Umat_k = [\phivec_{k_1},\phivec_{k_2},\dots,\phivec_{k_\nblk}] \in \realNum^{n\times \nblk}\label{eqn:weightU} \\ \label{eqn:batchevs}
 \text{ and }
\SigmaMat_k = \mathrm{diag}(\sigma_{k_1},\sigma_{k_2}\cdots,\sigma_{k_\nblk})  \in \realNum^{\nblk\times \nblk},
\end{eqnarray}
respectively. The weighting of the eigenvectors in equation (\ref{eqn:weightU}) guarantees orthonormality 
\begin{equation}
\PhiMat_k^*\Wmat\PhiMat_k = \Imat
\end{equation}
under the inner product (\ref{eqn:innerp}).

\section{Streaming SPOD}\label{onlinespod}

\begin{figure}
\centering
	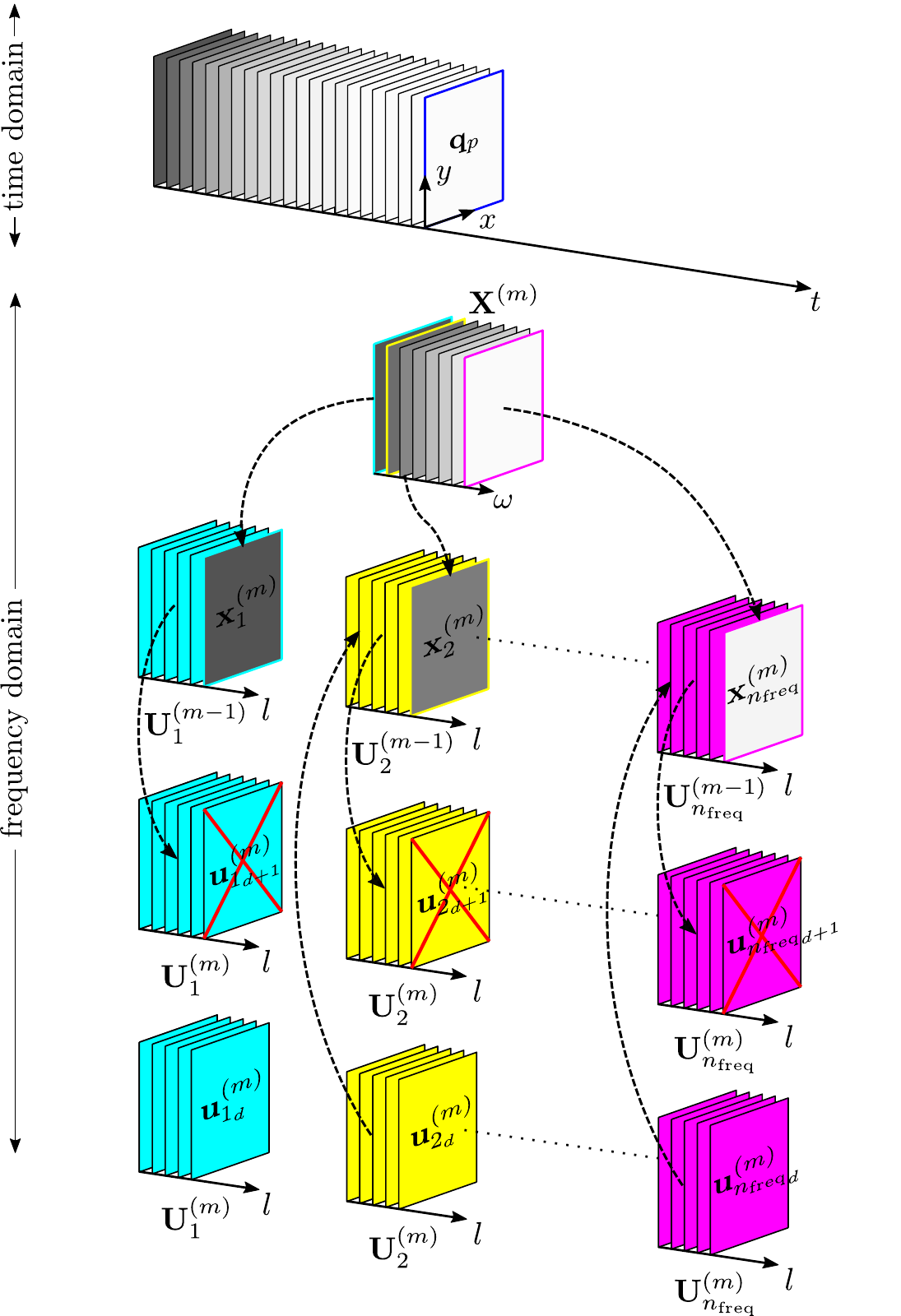	
	\caption{Illustration of the streaming SPOD algorithm. Numbers in parentheses denote the equations. As soon as a new data snapshot becomes available, the partial Fourier sums are updated. Once the Fourier sums are completed, the old eigenbases for each frequency are augmented by the orthogonal complement from the new data. The basis rotation and truncation conclude the update.}
	\label{fig:SPODsketch_online}
\end{figure}

Two aspects of the batch SPOD algorithm make it computationally demanding for large data sets.  First, $\nfreq$ snapshots must be loaded into memory and operated upon simultaneously in order to compute the required Fourier modes.  Second, $\nblk$ realizations of the Fourier mode at a given frequency of interest must be loaded into memory and operated upon simultaneously to compute the singular value decomposition that produces the SPOD modes.  In the following subsections, we develop strategies to overcome these two challenges, leading to a streaming algorithm that requires access to only the most recent data snapshot and recursively updates the $d$ most energetic SPOD modes for each frequency of interest.  A graphical illustration of the streaming algorithm is shown in figure \ref{fig:SPODsketch_online}.


\subsection{Streaming Fourier Sums}

Ideally, a streaming SPOD algorithm would require access to only one snapshot of the data at a time, e.g., the solution computed in a simulation or measured in an experiment at the most recent time instant.  The batch SPOD algorithm does not have this property because the discrete Fourier modes in equation~(\ref{eqn:dft_block}) are typically computed using the Fast Fourier Transform (FFT) algorithm, which requires simultaneous access to $n_{freg}$ snapshots.  This requirement can be relaxed by computing the Fourier modes using the original definition of the discrete Fourier transform rather than the FFT algorithm.  

Consider the definition of the discrete Fourier transform, 
\begin{equation}
\label{eqn:dft_def}
\hat{\qvec}_{k}^{(l)} = \sum \limits_{j=1}^{\nfreq} \qvec_{j}^{(l)} f_{jk},
\end{equation}
where
\begin{equation}
\label{eqn:dft_def_f}
f_{jk} = z^{(k-1)(j-1)}
\end{equation}
and $z = \exp( -\ii 2 \pi / \nfreq)$ is the primitive $\nfreq$-th root of unity.  Equation~(\ref{eqn:dft_def}) shows how each snapshot $\qvec_{j}^{(l)}$  contributes to each Fourier mode $\hat{\qvec}_{k}^{(l)}$ -- specifically, the snapshot at time $j$ is multiplied by the complex scalars $f_{jk}$ and then added to the contributions of other time instances to obtain each Fourier mode.  

This observation provides a way to compute the Fourier modes that requires access to only the most recent data snapshot.  A new snapshot $\qvec_{p}$ will appear in block $l$ if $1 < p - (l-1)(\nfreq - \novlp)<\nfreq$, in which case $\qvec_{j}^{(l)}$ is defined by equation~(\ref{eqn:snapshot_placement}) with $j = p - (l-1)(\nfreq - \novlp)$.  The snapshot $\qvec_{p}$ can appear in multiple blocks if the overlap $\novlp$ is nonzero.  Next, each $\qvec_{j}^{(l)}$ is multiplied by the corresponding $f_{jk}$ values for each $k = 1, \dots, \nfreq$ and added to previous terms to give the partial sum
\begin{equation}
\label{eqn:dft_partial}
\left[ \hat{\qvec}_{k}^{(l)} \right]_{n_p} = \left[ \hat{\qvec}_{k}^{(l)} \right]_{n_p-1} +  \qvec_{n_{p}}^{(l)} f_{n_{p}k} = \sum \limits_{j=1}^{n_{p}} \qvec_{j}^{(l)} f_{jk}.
\end{equation}
Once $n_p = \nfreq$ for block $l$, the Fourier mode is recovered as
\begin{equation}
\label{eqn:dft_partial_converge}
\hat{\qvec}_{k}^{(l)} =  \left[ \hat{\qvec}_{k}^{(l)} \right]_{\nfreq}.
\end{equation}

This procedure has several desirable properties.  First, by construction, it requires access to only the most recent data snapshot.  This immediately reduces the memory required to compute the Fourier modes by a factor of approximately $\nfreq$ compared to a standard FFT algorithm.  Second, no approximations have been made, so the computed Fourier modes are exact.  Third, additional computational and memory saving may be obtained by computing the partial sums in equation~(\ref{eqn:dft_partial}) only for frequncies of interest.  Often, the value of $\nfreq$ required to control spectral leakage and aliasing is larger than the number of frequencies actually needed for analysis. Standard FFT algorithms automatically compute every frequency, $k = 1, \dots, \nfreq$, whereas it is straightforward to compute only the frequencies of interest using the streaming algorithm by including only those values of $k$.  

The main drawback of the method is that computing all $\nfreq$ frequencies requires order $\nfreq^{2}$ operations, compared to $\nfreq \log \nfreq$ for an FFT algorithm.   However, memory requirements, not operation counts, are the primary obstacle for applying SPOD to large data sets.  Moreover, the increased operation count can be partially negated by computing only frequencies of interest, as described above.

\subsection{Incremental updates of the CSD}

The second aspect of batch SPOD that hinders its application to large data sets is the need to store many realizations of each Fourier mode in memory to compute the modes.  To overcome this obstacle, we develop an algorithm that recursively updates the $d$ most energetic SPOD modes for each frequency as new Fourier modes become available from the streaming Fourier algorithm. We require the updating algorithm to converge a fixed number of modes $d$ to be able to operate within a strictly limited amount of memory. We start by adapting Brand's \cite{brand2006fast} incremental SVD algorithm to the special case of updating the eigendecomposition of the estimated CSD matrix. The best rank-$d$ approximation used to truncate the eigenbasis and the initialization of the algorithm are discussed later in \S \ref{rankd} and \S \ref{init}, respectively. 

The block-wise sample mean is readily updated through the recursive relation
\begin{equation}
\qmean^{(m)} = \frac{m-1}{m}\qmean^{(m-1)} + \frac{1}{m}\left[\frac{1}{\nfreq}\sum_{j=1}^\nfreq \qvec_j^{(m)} \right].
\end{equation}
Analogously, a rank-1 update of the CSD takes the form
\begin{equation}\label{eqn:csdnew}
\Smat_k^{(m)} = \frac{m-1}{m}\Smat_k^{(m-1)} + \frac{m-1}{m^2}\xvec_k^{(m)}\xvec_k^{*(m)}\end{equation}
and can be performed once the $m$-th Fourier realization $\qfft_k^{(m)}$ becomes available.
Note that we use the sample CSD as an unbiased estimator for the unknown population CSD. The update formula for the CSD, equation (\ref{eqn:csdnew}), can be rewritten in terms of the data matrix $\Xmat_k$ as
\begin{equation} \label{xxupdate}
\Xmat_k^{(m)}\Xmat_k^{*(m)} 
= \frac{m-1}{m} \Xmat_k^{(m-1)}\Xmat_k^{*(m-1)} + \frac{m-1}{m^2}\xvec_k^{(m)}\xvec_k^{*(m)}
\end{equation}
by using definition (\ref{eqn:csd}). Analogous to equation (\ref{eqn:xmat}), we denote by 
\begin{equation}
\Xmat_k^{(m)}=[\xvec_k^{(1)},\xvec_k^{(2)},\dots,\xvec_k^{(m)}]=\frac{1}{\sqrt{m}}\Wmat^{\frac{1}{2}}\Qmatfft_k^{(m)} \in\realNum^{n\times m}
\end{equation}
the data matrix containing the first $m$ weighted Fourier realizations at the $k$-th frequency. We now insert the SVD of the data matrix 
\begin{equation}
\Xmat_k^{(m)}=\Umat_k^{(m)}\SigmaMat_k^{(m)}\Vmat_k^{*(m)}
\end{equation}
into the update formula (\ref{xxupdate}) to obtain an updating scheme 
\begin{equation}
\Umat_k^{(m)}\SigmaMat_k^{2(m)}\Umat_k^{*(m)} = \frac{m-1}{m} \Umat_k^{(m-1)}\SigmaMat_k^{2(m-1)}\Umat_k^{*(m-1)} + \frac{m-1}{m^2}\xvec_k^{(m)}\xvec_k^{*(m)}
\label{eq:us2u}
\end{equation}
for the eigendecomposition of the CSD at iteration level $m$ in terms of the eigendecomposition at level $m-1$ and the newly arrived data $\xvec_k^{(m)}$. For brevity, we factorize equation (\ref{eq:us2u}) and consider the data matrix
\begin{eqnarray}
\Xmat_k^{(m)}&=&
\begin{bmatrix}
\sqrt{\frac{m-1}{m}}\Umat_k^{(m-1)}\SigmaMat_k^{(m-1)}\Vmat_k^{*(m-1)} & \sqrt{\frac{m-1}{m^2}}\xvec_k^{(m)}
\end{bmatrix}
\label{eqn:xusvx}
\\ &=& 
\begin{bmatrix}
\Umat_k^{(m-1)} & \xvec_k^{(m)}
\end{bmatrix}
\begin{bmatrix}
\sqrt{\frac{m-1}{m}}\SigmaMat_k^{(m-1)}\Vmat_k^{*(m-1)} & \Zmat \\ \Zmat & \sqrt{\frac{m-1}{m^2}}
\end{bmatrix}
\label{eqn:uxmat}
\end{eqnarray}
instead of the product $\Xmat_k^{(m)}\Xmat_k^{*(m)}$. With the goal in mind to update $\Umat_k^{(m-1)}$ with the new data $\xvec_k^{(m)}$, equation (\ref{eqn:xusvx}) is factored into the matrix product given by equation (\ref{eqn:uxmat}). We seek to find the updated set of left singular vectors $\Umat_k^{(m)}$ in the column space of the augmented eigenbasis $\left[\Umat_k^{(m-1)}\quad \xvec_k^{(m)}\right]$ and start by restoring orthonormality. The component of $\xvec_k^{(m)}$ that is orthogonal to $\Umat_k^{(m-1)}$ can readily be found from a partial step of the modified Gram-Schmidt (MGS) algorithm as
\begin{equation}\label{eqn:mgs}
\uvec_k^{\perp(m)} = \xvec_k^{(m)} - \Umat_k^{(m-1)}\Umat_k^{*(m-1)}\xvec_k^{(m)}.\end{equation}
Using equation (\ref{eqn:mgs}), the multiplicand is recast into a product of a modified multiplicand $\left[\Umat_k^{(m-1)}\quad \frac{\uvec_k^{\perp(m)}}{\|\uvec_k^{\perp(m)}\|}\right]$ with orthonormal columns and a matrix as
\begin{equation}\label{eqn:uu}
\begin{bmatrix}
\Umat_k^{(m-1)} & \xvec_k^{(m)}
\end{bmatrix}
=
\begin{bmatrix}
\Umat_k^{(m-1)} & \frac{\uvec_k^{\perp(m)}}{\|\uvec_k^{\perp(m)}\|}
\end{bmatrix}
\begin{bmatrix}
\Imat & \Umat_k^{*(m-1)}\xvec_k^{(m)} \\ \Zmat & \|\uvec_k^{\perp(m)}\|
\end{bmatrix}.
\end{equation}
Inserting equation (\ref{eqn:uu}) into equation (\ref{eqn:uxmat}) yields the expression 
\begin{multline}\label{eqn:xnew}
\Xmat_k^{(m)} = 
\begin{bmatrix}
\Umat_k^{(m-1)} & \frac{\uvec_k^{\perp(m)}}{\|\uvec_k^{\perp(m)}\|}
\end{bmatrix} \\ \times
\begin{bmatrix}
\Imat & \Umat_k^{*(m-1)}\xvec_k^{(m)} \\ \Zmat & \|\uvec_k^{\perp(m)}\|
\end{bmatrix}
\begin{bmatrix}
\sqrt{\frac{m-1}{m}}\SigmaMat_k^{(m-1)}\Vmat_k^{(m-1)} & \Zmat \\ \Zmat & \sqrt{\frac{m-1}{m^2}}
\end{bmatrix}
\end{multline}
for the updated data matrix. Multiplying equation (\ref{eqn:xnew}) with its conjugate transpose yields the updated CSD 
\begin{equation}\label{eqn:csdnewx}
\Xmat_k^{(m)}\Xmat_k^{*(m)} = 
\begin{bmatrix}
\Umat_k^{(m-1)} & \frac{\uvec_k^{\perp(m)}}{\|\uvec_k^{\perp(m)}\|}
\end{bmatrix}
\Mmat
\begin{bmatrix}
\Umat_k^{*(m-1)} \\ \frac{\uvec_k^{\perp*(m)}}{\|\uvec_k^{\perp(m)}\|}
\end{bmatrix},
\end{equation}
where
\begin{equation}
\Mmat=
{\frac{m-1}{m^2}}
\begin{bmatrix}\small
{m}\SigmaMat_k^{(m-1)^2}+\Umat_k^{*(m-1)}\xvec_k^{(m)}\xvec_k^{*(m)}\Umat_k^{(m-1)} & \|\uvec_k^{\perp(m)}\|\Umat_k^{*(m-1)}\xvec_k^{(m)} \\ \|\uvec_k^{\perp(m)}\|\xvec_k^{*(m)}\Umat_k^{(m-1)} & \|\uvec_k^{\perp(m)}\|^2
\end{bmatrix}
\end{equation}
is a $m\times m$ Hermitian matrix. The remaining task is to recast the right-hand side of equation (\ref{eqn:csdnewx}) into SVD form. This is achieved through an eigendecomposition $\Mmat=\tilde{\Umat}\tilde{\SigmaMat}^2\tilde{\Umat}^*$ of $\Mmat$. Equivalently, we may factor $\Mmat$ as $\Mmat=\Kmat\Kmat^*$ first, where
\begin{equation}\label{eqn:kmat}
\Kmat=
\sqrt{\frac{m-1}{m^2}}
\begin{bmatrix}
\sqrt{m}\SigmaMat_k^{(m-1)} & \Umat_k^{*(m-1)}\xvec_k^{(m)} \\ \Zmat & \|\uvec_k^{\perp(m)}\|
\end{bmatrix} =\tilde{\Umat}\tilde{\SigmaMat}\tilde{\Vmat}^*,
\end{equation}
and compute the SVD of $\Kmat$. Inserting the eigendecomposition $\Mmat=\Kmat\Kmat^*=\tilde{\Umat}\tilde{\SigmaMat}^2\tilde{\Umat}^*$ into equation (\ref{eqn:csdnewx}) yields
\begin{multline}\label{eqn:evdnew}
\Xmat_k^{(m)}\Xmat_k^{*(m)} =
\underbrace{
\begin{bmatrix}
\Umat_k^{(m-1)} & \frac{\uvec_k^{\perp(m)}}{\|\uvec_k^{\perp(m)}\|}
\end{bmatrix}
\tilde{\Umat}
}_{\Umat_k^{(m)}}
\underbrace{\tilde{\SigmaMat}}_{\SigmaMat_k^{(m)}}
\tilde{\SigmaMat}^*\tilde{\Umat}^*
\begin{bmatrix}
\Umat_k^{*(m-1)} \\ \frac{\uvec_k^{\perp*(m)}}{\|\uvec_k^{\perp(m)}\|}
\end{bmatrix} \\
= \Umat_k^{(m)}\SigmaMat_k^{(m)^2}\Umat_k^{*(m)}.
\end{multline}
By noting that $\tilde{\Umat}$ is a rotation matrix that preserves orthonormality, the spectral theorem guarantees that this decomposition is unique, and therefore corresponds to the updated eigendecomposition of the CSD matrix. The updates of the eigenbasis and eigenvalues hence take the form
\begin{eqnarray}
\label{eqn:rot}
{\Umat_k^{(m)}}&=&\begin{bmatrix}
\Umat_k^{(m-1)} \quad \frac{\uvec_k^{\perp(m)}}{\|\uvec_k^{\perp(m)}\|}
\end{bmatrix}
\tilde{\Umat}
\quad\text{ and }\\
{\SigmaMat_k^{(m)}} &=& \tilde{\SigmaMat},
\end{eqnarray}
respectively. Besides the rotation (\ref{eqn:rot}), the implementation of the algorithm requires the MGS step (\ref{eqn:mgs}) and the construction and SVD of $\Kmat$, as defined in equation (\ref{eqn:kmat}). Note that the large matricies $\Xmat_k^{(m)}\Xmat_k^{*(m)}$ and $\Umat_k^{(m-1)}\Umat_k^{*(m-1)}$ appearing in the derivation are never computed in the actual algorithm. Up to this point, no approximations have been made.

\subsection{Eigenbasis truncation}\label{rankd}
The recursive rank-1 updates described by equation (\ref{eqn:rot}) add an additional vector to the eigenbasis of the CSD matrix at each step. In practice, however, we are interested in converging a fixed number $d$ of the most energetic SPOD modes only. Fortunately, the basic property of the SVD guarantees that this best rank-$d$ approximation is readily obtained by truncating the basis after the $d$-th vecotor. Formally, we express this by partitioning the updated eigenbasis and matrix of singular values as 
\begin{equation}\label{eqn:rank}
{\Umat_k^{(m)}} = 
\begin{bmatrix}
{\Umat_k'}^{(m)} & \uvec_{k_{d+1}}
\end{bmatrix}
\text{ and }
\SigmaMat_k^{(m)} =
\begin{bmatrix}
{{\SigmaMat}'_k}^{(m)} & \Zmat \\ \Zmat & {\sigma}_{k_{d+1}}
\end{bmatrix},
\end{equation}
respectively, and letting 
\begin{equation}
\Umat_k^{(m)} \gets {\Umat_k'}^{(m)} \text{ and } \SigmaMat_k^{(m)} \gets {\SigmaMat'}^{(m)} \label{eqn:update}
\end{equation}
as we update the basis during runtime. At this point, a truncation error is introduced as the vector component $\uvec_{k_{d+1}}$ that is orthogonal to the retained $d$ eigenvectors is discarded. The batch SPOD algorithm, on the contrary, guarantees that every eigenvector is orthogonal to all other $\nblk-1$ eigenvectors. We address the error resulting from the basis truncation in \S \ref{errors}.

As before, the final step of the algorithm consists of obtaining the SPOD modes by weighting the CSD eigenvectors according to $\PhiMat_k^{(m)}=\Wmat^{-\frac{1}{2}}\Umat_k^{(m)} = [\phivec_{k_1}^{(m)},\phivec_{k_2}^{(m)},\dots,\phivec_{k_d}^{(m)}] \in \realNum^{n\times d}$.

\subsection{Initialization}\label{init}
 Once the first Fourier realization becomes available, the eigenbasis is initialized as $\Umat_k^{(1)} \gets [\xvec_{k}^{(1)},\Zmat,\dots,\Zmat] \in \realNum^{n\times d}$ and subsequenly updated as $\Umat_k^{(2)} = [\uvec_{k_1}^{(2)},\uvec_{k_2}^{(2)},\Zmat,\dots,\Zmat]$ at iteration level $m=2$, and so on. Correspondingly, the singular value matrix is initialized with the first Fourier realization as $\SigmaMat_k^{(1)} \gets \mathrm{diag}(\xvec_{k}^{*(1)}\xvec_{k}^{(1)},0,\dots,0) \in \realNum^{d\times d}$ before being updated as $\SigmaMat_k^{(2)} = \mathrm{diag}(\sigma_{k_1}^{(2)},\sigma_{k_2}^{(2)},0,\dots,0)$. The truncation of the eigenbasis is performed once the iteration level exceeds the number of desired SPOD modes, i.e.~when $m\geq d+1$.
 
Alternatively, the eigenbasis can be initialized from a previously computed SPOD basis as 
$\Umat_k^{(1)}=\Wmat^{\frac{1}{2}}\PhiMat_k^\text{old}$. Initializing the algorithm with an initial SPOD basis $\PhiMat_k^\text{old}$ obtained from a batch computation or a streaming computation with a larger value $d$ has the benefit of reducing the truncation error. This follows directly from the best rank-$d$ property of the SVD.

\section{Errors and convergence}\label{errors}
The errors of the approximation can be quantified by comparing the rank-$d$ solutions at the $m$-th iteration level to the reference solution $\PhiMat_k^\text{batch}$ and $\SigmaMat_k^\text{batch}$ obtained from the batch algorithm described in \S \ref{batch}.

\paragraph{Errors with respect to batch solution} We define two error quantities
\begin{eqnarray}
e^{\phi,\text{batch}}_{j}(m) &=& \sum_{k=1}^{\nfreq}\left( 1-\underset{j}{\max}\left<\phivec^{(m)}_{k_j},\phivec_{k_j}^\text{batch}\right>_E\right) \text{ and} \label{eq:error_mode_batch}
\\ 
e^{\lambda,\text{batch}}_{j}(m) &=& \sum_{k=1}^{\nfreq}\left|\frac{\lambda_{k_j}^{(m)}-\lambda_{k_j}^\text{batch}}{\lambda_{k_j}^\text{batch}}\right|.  \label{eq:error_energy_batch}
\end{eqnarray}
that measure the error in the $j$-th eigenvector and eigenvalue, respectively. The eigenvector error given by equation (\ref{eq:error_mode_batch}) is defined in terms of the inner product (\ref{eqn:innerp}) and compares the patterns of two modes, i.e.~it is 0 for identical and 1 for orthogonal modes. The maximum over the mode rank index is taken to ensure that the most similar modes are compared to each other. This is important as similar modes can swap order between iterations. 

\paragraph{Convergence with respect to previous solution} Estimates for the convergence of the eigenvectors and eigenvalues are defined analogously in terms of their values at the previous iteration level $m-1$. The resulting measures of convergence
\begin{eqnarray}
e^{\phi,\text{prev}}_{j}(m) &=& \sum_{k=1}^{\nfreq}\left(1-\underset{j}{\max}\left<\phivec^{(m)}_{k_j},\phivec^{(m-1)}_{k_j}\right>_E \right) \text{ and} \label{eq:error_mode_prev}
\\
e^{\lambda,\text{prev}}_{j}(m) &=& \sum_{k=1}^{\nfreq}\left|\frac{\lambda_{k_j}^{(m)}-\lambda_{k_j}^{(m-1)}}{\lambda_{k_j}^{(m-1)}}\right|,  \label{eq:error_energy_prev}
\end{eqnarray}
for the $j$-th eigenfunction and eigenvalue, respectively, can be monitored during runtime.

\section{Examples}\label{examples}
This section demonstrates the performance of the proposed streaming SPOD
algorithm on two examples. The first example is a high-fidelity numerical simulation of a turbulent jet \cite{bresetal_2018jfm}, and the second example is optical flow obtained from a high-speed movie of a stepped spillway experiment \cite{Kramer2017,Zhang2017}. An overview of the databases and SPOD parameters is presented in table \ref{tab:ex}. The SPOD parameters are chosen according to best practice. A discussion of how to choose them is beyond the scope of this work. The same applies to detailed physical interpretations of the results. Here, we focus on the performance and convergence of the streaming algorithm as compared to its offline batch counterpart. We use a Hanning window for the Fourier transformation and set the number of retained SPOD modes to $d=5$. The effect of eigenbasis truncation is discussed in more detail in \S \ref{rankderror}.

\begin{table}
  \begin{center}
\def~{\hphantom{0}}
  \begin{tabular}{lcccc|ccc|c}
  \multicolumn{5}{ c }{Database}  & \multicolumn{4}{ c }{SPOD parameters} \\ 
      Case & $\qvec$ & $n_{x_1}$ & $n_{x_2}$ & $n_{t}$ & $n_{\text{freq}}$& $n_{\text{ovlp}}$& $n_{\text{blk}}$ & $d$  \\[3pt]\hline
      Jet      & $p^{\mathrm{symm}}$ & 175 & 39 & 10,000 & 256 & 128 & 78 & 5 \\
      Spillway & $[u,v]$ & 224 & 160 & 19,782 & 512 & 256 & 77 & 5
  \end{tabular}
  \caption{Parameters for the two example databases and the SPOD. The spectral estimation parameters $n_{\text{freq}}$, $n_{\text{ovlp}}$ and $n_{\text{blk}}$ are identical for batch and streaming SPOD. $d$ is the number of desired modes for the streaming algorithm.}
  \label{tab:ex}
  \end{center}
\end{table}

\subsection{Example 1: large eddy simulation of a turbulent jet}\label{jet}

The turbulent jet is a typical examples of a stationary flow. A number of studies, see e.g.~\cite{glauser1987coherent} for an early experimental and \cite{SchmidtEtAl_2017_JFM} for a recent numerical example, use SPOD to analyze jet turbulence. Our first is example is an LES of a Mach 0.9 jet at a jet diameter-based Reynolds number of $1.01\cdot10^6$ \cite{bresetal_2018jfm}. The LES was calculated using the unstructured flow solver ``Charles" \cite{bres2017unstructured}. The database consists of 10,000 snapshots of the axisymmetric component of the pressure field obtained as the zeroth azimuthal Fourier component of the flow. We choose to resolve 129 positive frequencies by setting $n_{\text{freq}}=256$. Each block therefore consists of 256 snapshots. We further use a 50\% overlap by letting $n_{\text{ovlp}}=128$. This results in a total of 77 blocks for the spectral estimation, each of which represents one realization of the temporal Fourier transform.
\begin{figure}
\centering
	\includegraphics[trim=0 0cm 0 0mm, clip, width=1\textwidth]{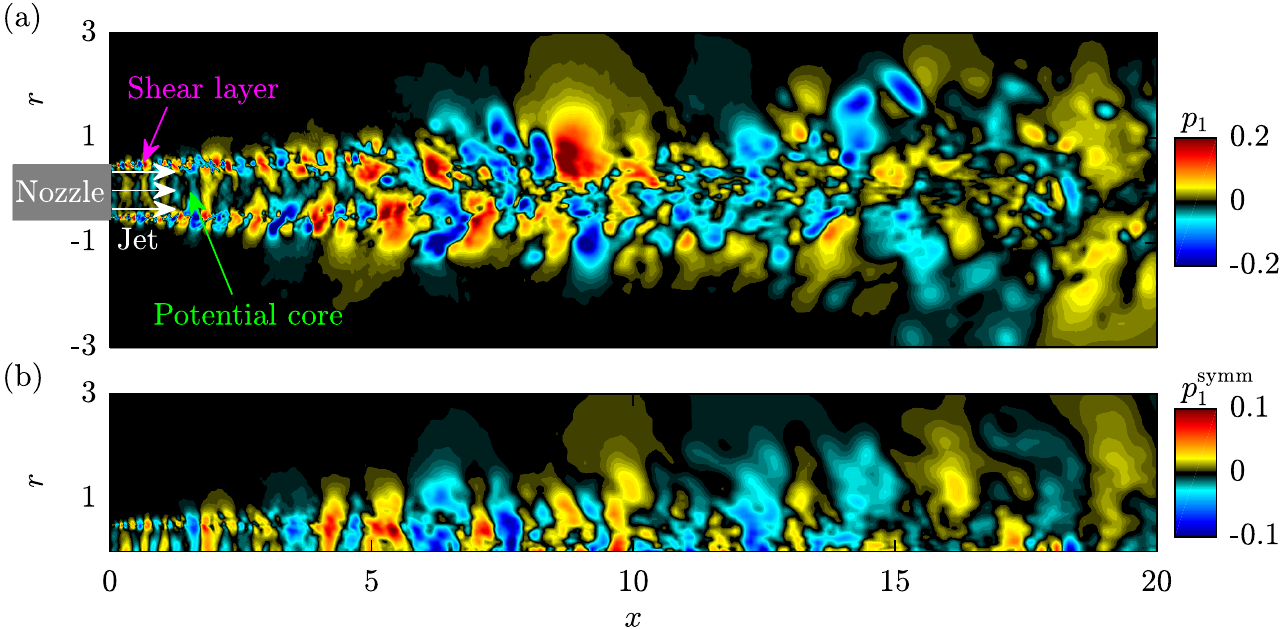}
	\caption{Fluctuating pressure of the first snapshot of the turbulent jet LES: (a) streamwise plane; (b) symmetric pressure component only.  The boundary layer inside the nozzle is turbulent, whereas the flow inside the potential core is laminar. The potential core collapses after approximately 5 jet diameters.}
	\label{fig:jet_example}
\end{figure}
The first snapshot of the database is visualized in figure \ref{fig:jet_example}. The chaotic nature of the flow becomes apparent at first glance.

\begin{figure}
\centering
	\includegraphics[trim=0 0cm 0 0mm, clip, width=1\textwidth]{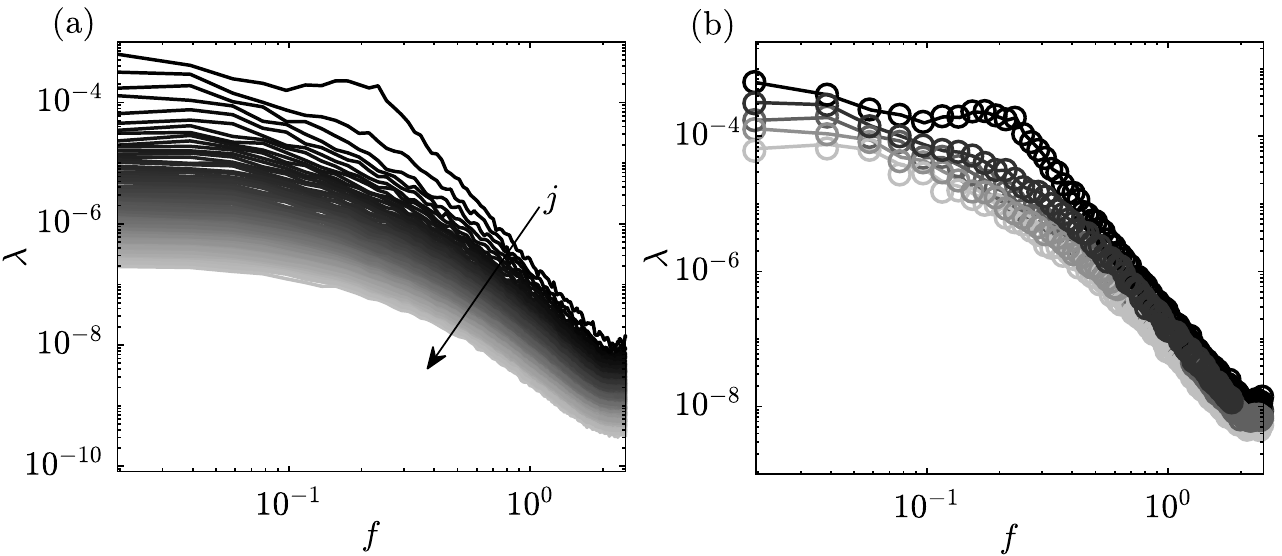}
	\caption{SPOD energy spectra of the turbulent jet obtained using batch SPOD and streaming SPOD: (a) all $\nblk=77$ eigenvalues computed using batch SPOD ($-\!\!\!-\!\!\!-$); (b) $d=5$ leading eigenvalues calculated using streaming SPOD ($\circ$). The batch solution ($-\!\!\!-\!\!\!-$) is shown for comparison. $j$ indicates the mode index from black ($j=1$, most energetic) to light gray ($j=\nblk$ in (a) and $j=d$ in (b), least energetic).}
	\label{fig:jet_spectrum}
\end{figure}
Figure \ref{fig:jet_spectrum}(a) shows the batch SPOD  spectrum obtained for the spectral estimation parameters listed in table \ref{tab:ex}. Each line represents the energy spectrum associated with a single mode index $j$. The total number of modes is equal to the number of blocks, i.e.~$\nblk=77$ in this example. Most of the energy is concentrated in the large-scale structures that dominate at low frequencies. The roll-off of the distribution at higher frequencies is indicative of an energy cascade that transfers energy from larger to smaller scales. Over a certain frequency interval $0.1\lesssim f \lesssim 0.6$, the first mode is significantly more energetic that the other modes. This low-rank behavior has important physical implications discussed elsewhere \cite{SchmidtEtAl_2018_JFM}. The spectra of the five leading modes are replicated in figure \ref{fig:jet_spectrum}(b) and compared to the results obtained using streaming SPOD ($\circ$ symbols). It can be seen that the two results are almost indistinguishable. This provides a first indication that the streaming SPOD algorithm accurately approximates the SPOD eigenvalues. We will quantify this observation in the context of figure \ref{fig:jet_error} below.

\begin{figure}
\centering
	\includegraphics[trim=0 2cm 0 0mm, clip, width=1\textwidth]{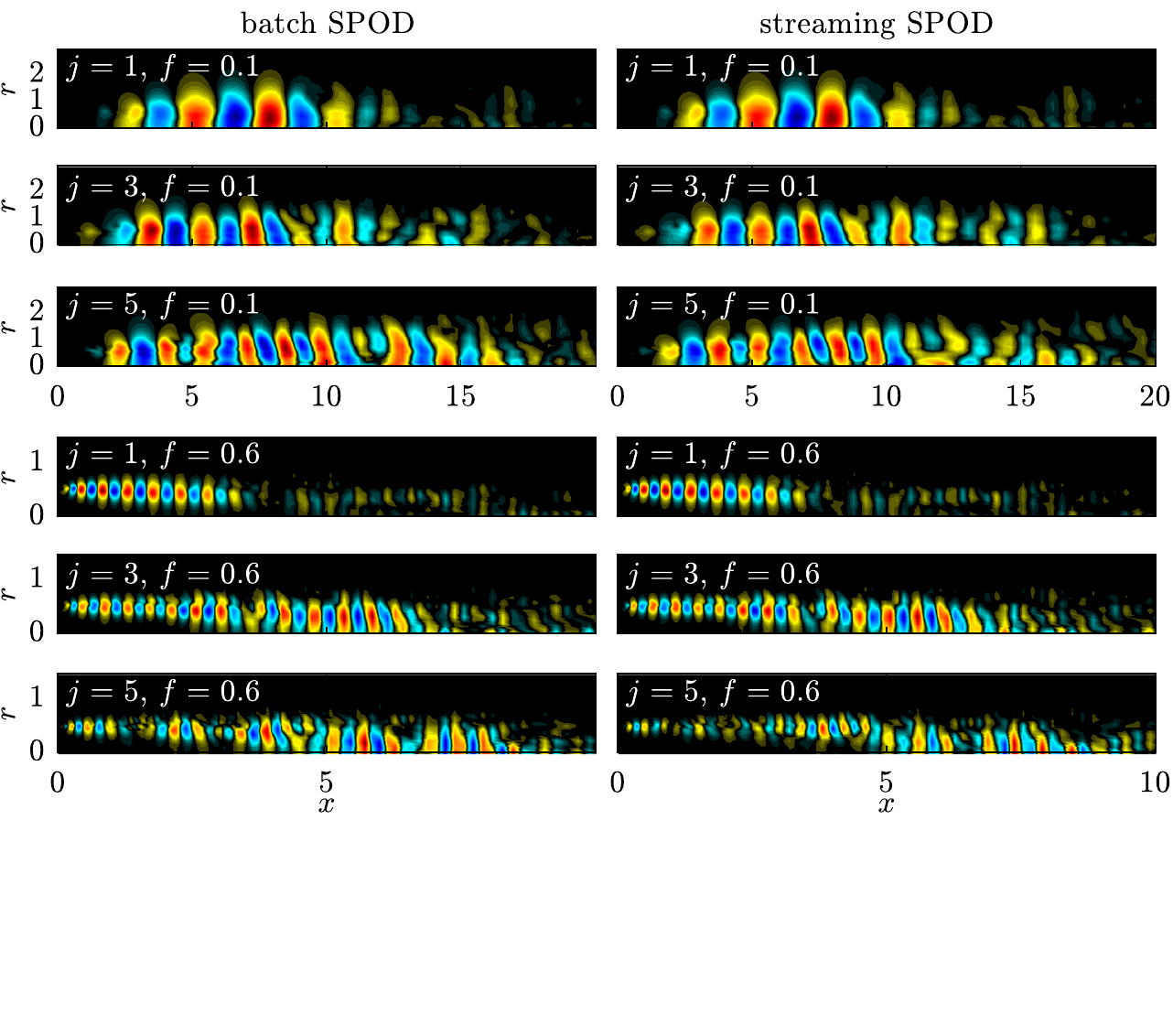}
	\caption{Side-by-side comparison of SPOD modes of the pressure field calculated using batch SPOD (left column) and streaming SPOD (right column) for the jet example.}
	\label{fig:jet_compare_modes}
\end{figure}
After establishing that the modal energies are well approximated by the streaming algorithm, we next examine the modal structures. Figure \ref{fig:jet_compare_modes} shows a side-by-side comparison of the first ($j=1$), third ($j=3$) and fifth ($j=5$) modes at two representative frequencies ($f=0.1$, top half and $f=0.6$, bottom half). The leading modes (first and fourth row) computed using streaming SPOD are almost indistinguishable from the reference solution for both frequencies. The third modes (second and fifth row) still compare well. More differences become apparent for the fifth modes. It has to be kept in mind though, that the subdominant modes are in general more difficult to converge. This exemplifies the importance of being able to converge second-order statistics from long data sequences.

\begin{figure}
\centering
	\includegraphics[trim=0 0cm 0 0mm, clip, width=1\textwidth]{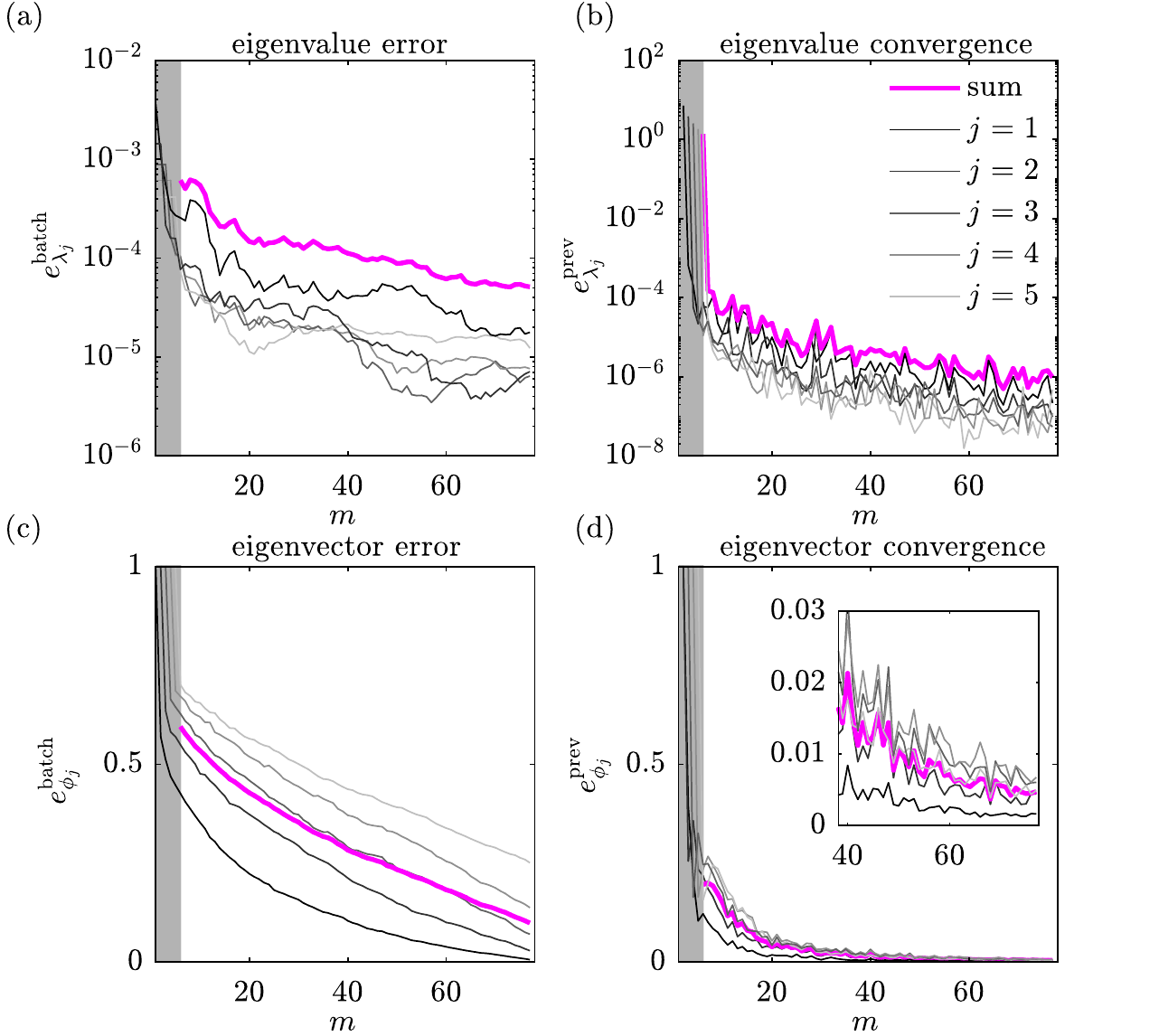}
	\caption{Streaming SPOD error and convergence for the turbulent jet: (a) eigenvalue error; (b) eigenvalue convergence; (c) eigenvector error; (d) eigenvalue convergence. The magenta lines show the cumulative error in (a,c) and the mean of the convergence metric in (b,d), respectively. The shaded area demarcates the initial region $1\leq m \leq d+1$ in which the eigenbasis is still rank deficient.}
	\label{fig:jet_error}
\end{figure}
In figure \ref{fig:jet_error}, we next investigate the errors and convergence behavior for the jet example in terms of the quantities defined in equations (\ref{eq:error_mode_batch})-(\ref{eq:error_energy_prev}). The eigenvalue error in figure \ref{fig:jet_error}(a) drops by approximately one order of magnitude from beginning to end. As anticipated from figure \ref{fig:jet_spectrum}(b), the eigenvalue error is generally small, i.e.~below the per mil range after the first iteration. The eigenvalue convergence is addressed in figure \ref{fig:jet_error}(b). Staring from the end of the initialization phase (gray shaded area), the convergence measure drops by about two orders of magnitude. The error and convergence of the eigenvectors are investigate in figure \ref{fig:jet_error}(c) and \ref{fig:jet_error}(d), respectively. Is is observed that the eigenvector error drops monotonically for all five modes. The similarity of the leading batch and streaming SPOD modes previously seen in figure \ref{fig:jet_compare_modes} is reflected by the small error of $0.6\%$. Similarly, the differences in the fifth modes result in a $25\%$ error according to the metric. Since the eigenvalue is accurately predicted at the same time, we conclude that this large error is primarily a result of the slow statistical convergence of the subdominant modes. The inset in figure \ref{fig:jet_error}(d) exemplifies this slow convergence. After about 50 iterations, the errors of most modes are below $1\%$.

\subsection{Example 2: optical flow of a stepped spillway}\label{spillway}
The second example is that of a laboratory stepped spillway \cite{Kramer2017}. Stepped spillways are hydraulic structures designed to control flow release and to achieve high energy dissipation. The two-phase flow of the laboratory spillway is filmed using a high-speed camera and an optical flow algorithm \cite{Liu2015,Zhang2017} was used to estimate the streamwise and normal velocity components of the air-water mixture. The parameters of the optical flow database are summarized in table \ref{tab:ex}. 

\begin{figure}
\centering
	\includegraphics[trim=0 0cm 0 0mm, clip, width=1\textwidth]{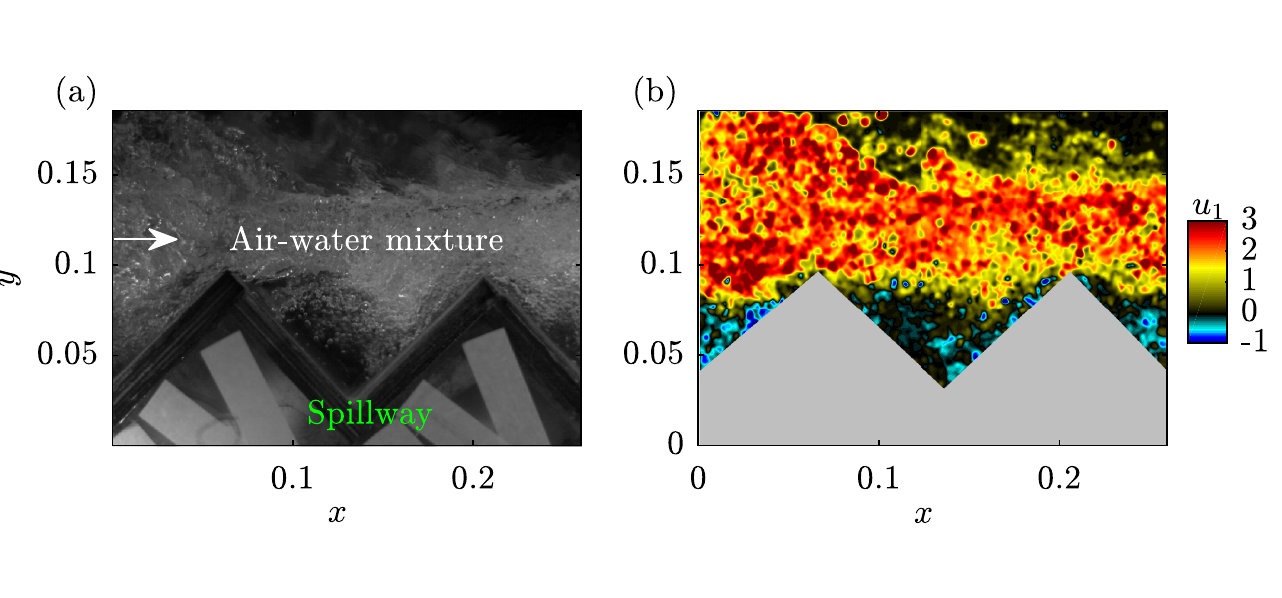}
	\caption{First snapshot of the stepped spillway: (a) raw video frame; (b) streamwise velocity computed using optical flow. The air-water flow is characterized by instability growth, air entrainment and strong turbulence.
	}
	\label{fig:spillway_example}
\end{figure}
Figure \ref{fig:spillway_example} shows an example of the raw video data and a processed snapshot of the instantaneous streamwise velocity component. As for the jet example, we will not address the complex multi-phase physics of the setup, but focus on the performance of the streaming SPOD algorithm instead. We have selected the spillway as a second example to investigate the algorithm's performance under high noise conditions. The high noise level of the measurement is apparent in figure \ref{fig:spillway_example}(b).

\begin{figure}
\centering
	\includegraphics[trim=0 0cm 0 0mm, clip, width=1\textwidth]{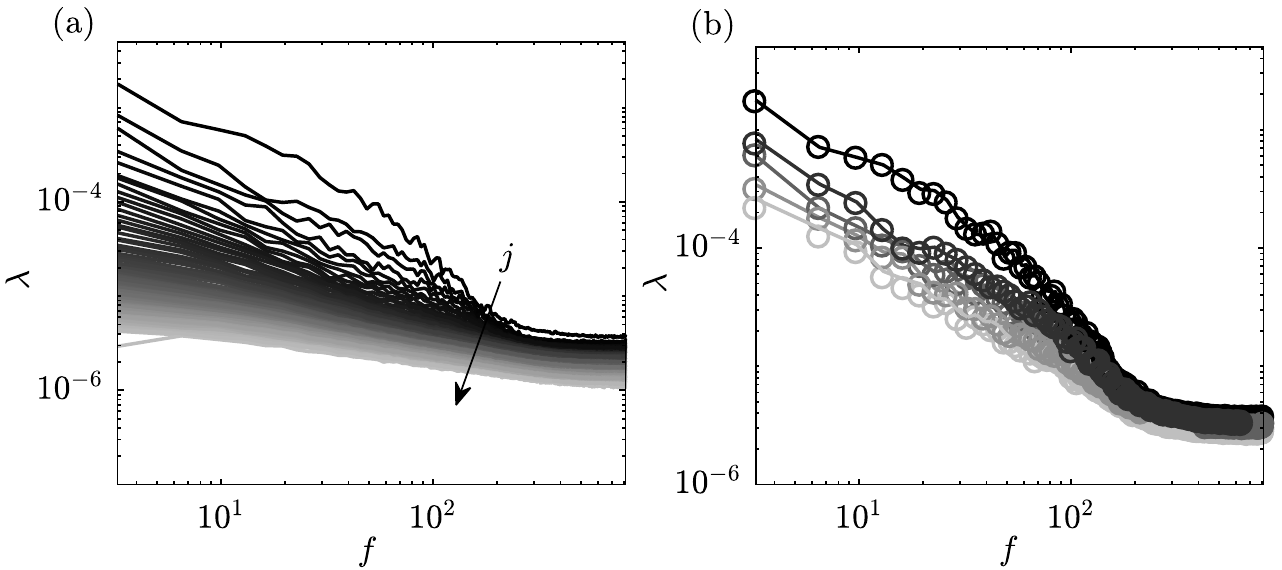}
	\caption{Same as figure \ref{fig:jet_spectrum}, but for the spillway example.}
	\label{fig:spillway_spectrum}
\end{figure}
As before, spectra obtained using batch SPOD and the streaming version are compared in figure \ref{fig:spillway_spectrum}. It is observed that the modal energies asymptote towards a constant value for $f\gtrsim200$. The plateau seen at these frequencies indicates the noise floor of the measurement. An inspection of the SPOD modes confirms this conjecture. Modes in this region are dominated by noise and show no apparent structure (not shown).

\begin{figure}
\centering
	\includegraphics[trim=0 2cm 0 0mm, clip, width=1\textwidth]{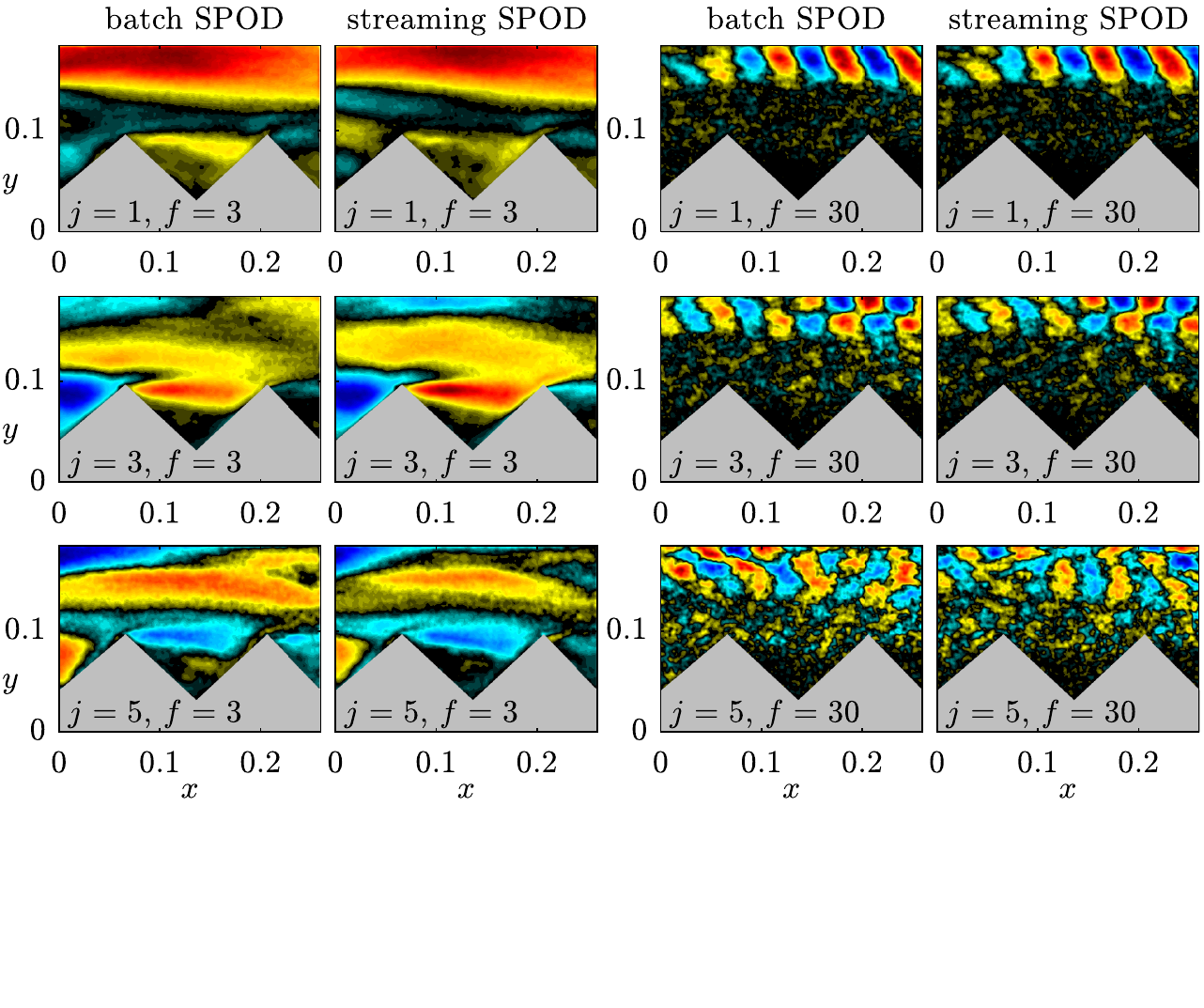}
	\caption{Side-by-side comparison of SPOD modes calculated using batch SPOD (first and third column) and streaming SPOD (second and fourth column) for the spillway example. The streamwise velocity is shown.}
	\label{fig:spillway_compare_modes}
\end{figure}
The comparison in figure \ref{fig:spillway_compare_modes} shows that the SPOD modes computed using the streaming algorithm closely resemble their batch SPOD counterparts. At the lower frequency (left), the SPOD modes are comprised of surface waves and oscillations of the shear-layer between the step ridges. Surface waves are the dominant structures at higher frequencies (right). Increasingly high noise levels are observed in the less energetic modes, in particular for the higher frequency case.

\begin{figure}
\centering
	\includegraphics[trim=0 0cm 0 0mm, clip, width=1\textwidth]{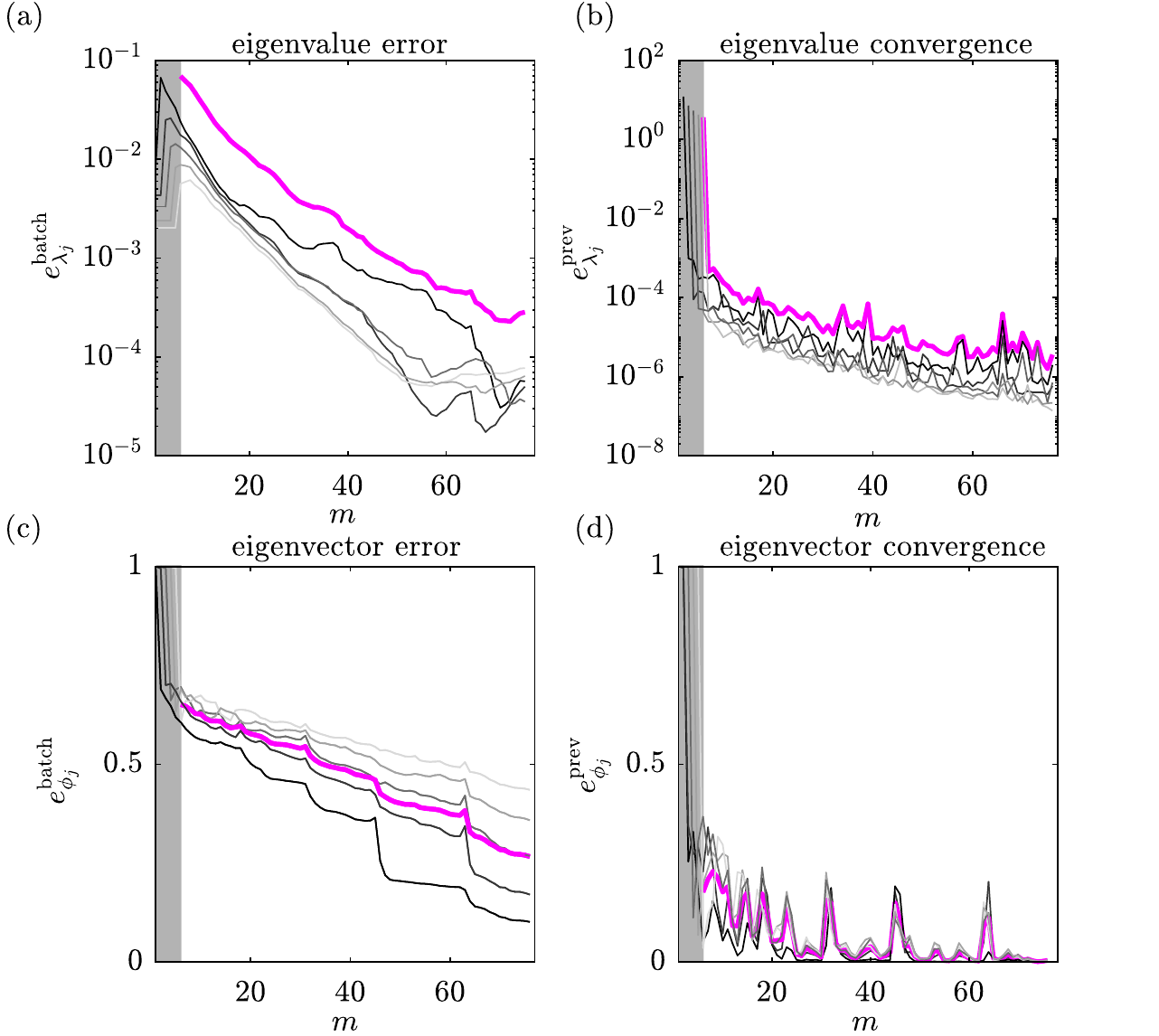}
	\caption{Same as figure \ref{fig:jet_error} but for the spillway example.}
	\label{fig:spillway_error}
\end{figure}
The eigenvalue error is studied in figure \ref{fig:spillway_error}(a). Initially, the error is significantly larger as compared to the turbulent jet case shown in figure \ref{fig:jet_error}(a). Subsequently, a faster drop-off allows the eigenvalue error to recover values similar to those found for the jet example. The eigenvalue convergence behavior shown in figure \ref{fig:spillway_error}(b) is very similar to that of the jet example. 

The eigenvector error and convergence are plotted in figure \ref{fig:spillway_error}(c) and \ref{fig:spillway_error}(d), respectively. Both metrics occasionally undergo rapid changes, most prominently at iteration level $m=45$. Sudden drops in the error are directly associated with peaks in the convergence measure. This behavior occurs when an eigenvector in the truncated basis gets replaced by a different structure. The re-orthogonalization of the eigenbasis after such an event leads a better correspondence with the batch solution. At the same time, the convergence measure spikes as a result of the change in modal structure. The error ranges between $10\%$ (first mode) and $44\%$ (fifth mode). The similarity of the modes depicted in figure \ref{fig:spillway_compare_modes} and the lower errors in the jet case, as seen in figure \ref{fig:jet_error}, strongly suggest that these relatively high errors are mainly associated with measurement noise. 

\section{Effect of eigenbasis truncation}\label{rankderror}
Spectral estimation parameters aside, the desired number of SPOD modes $d$ is the only additional user input required by the streaming algorithm. The basis truncation inevitably leads to an approximation error that originates from discarding the vector component orthogonal to the span of the existing basis vectors $\Umat_k^{(m-1)}$. 

\begin{figure}
\centering
	\includegraphics[trim=0 0cm 0 0mm, clip, width=1\textwidth]{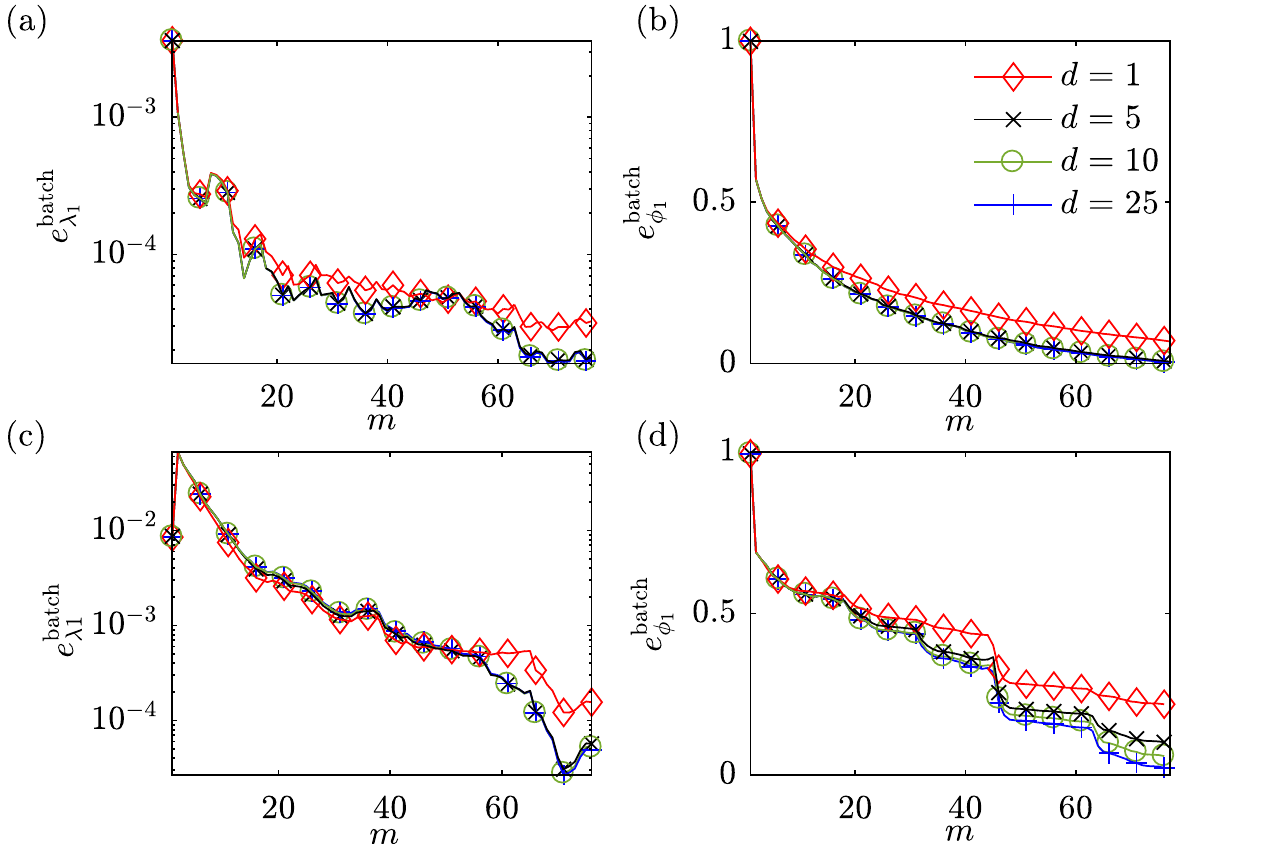}
	\caption{Eigenvalue (left column) and eigenvector (right column) errors of the first SPOD mode for different numbers of basis vectors $d$: (a,b) turbulent jet; (c,d) spillway.}
	\label{fig:rankd_error}
\end{figure}
Figure \ref{fig:rankd_error} compares eigenvalue and eigenvector errors of the first SPOD mode for four different values of $d$. The jet and spillway examples are shown in figure \ref{fig:rankd_error}(a,b) and \ref{fig:rankd_error}(c,d), respectively. It is observed that even restricting the basis to a single vector, i.e.~the most aggressive truncation possible, does not lead to significant errors. For $d\geq5$, all error metrics shown in figure \ref{fig:rankd_error}(a-c) are almost indistinguishable. Small differences are observed in the eigenvector error for the spillway example. In \ref{fig:rankd_error}(d), the eigenvector error ranges between $10\%$ for $d=5$ and $2\%$ for $d=20$. As discussed in figure \ref{fig:spillway_compare_modes}, this discrepancy is mainly related to data noise.

\begin{figure}
\centering
	\includegraphics[trim=0 0cm 0 0mm, clip, width=1\textwidth]{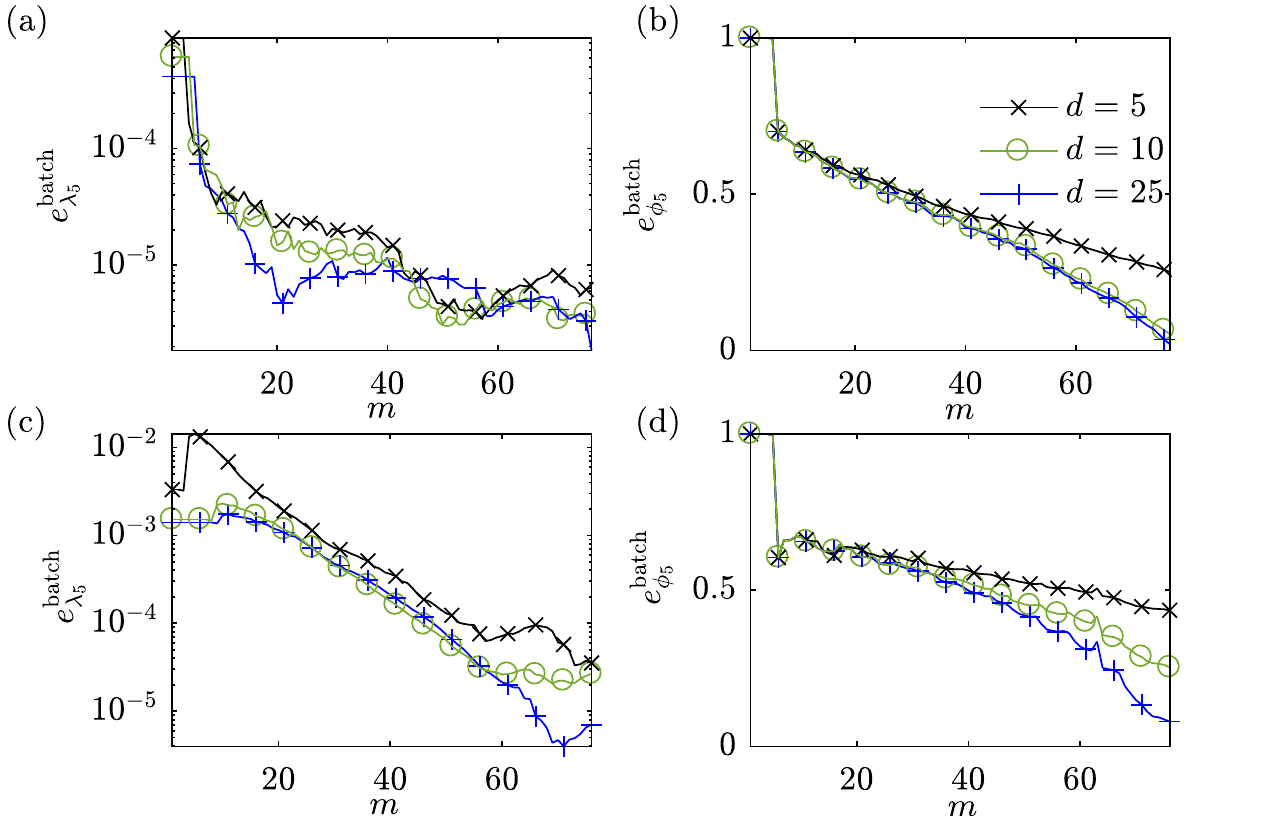}
	\caption{Eigenvalue (left column) and eigenvector (right column) errors of the fifth SPOD mode for different numbers of basis vectors $d$: (a,b) turbulent jet; (c,d) spillway.}
	\label{fig:rankd_error_mode5}
\end{figure}
After establishing that retaining only a few eigenvectors is sufficient to control the truncation error, we now focus on the effect of truncation on the suboptimal modes. Analogous to figure \ref{fig:rankd_error}, we compare the truncation errors of the fifth mode in figure \ref{fig:rankd_error_mode5}. Its error characteristics are similar to the ones of the first mode. This can be seen by comparing figure \ref{fig:rankd_error_mode5}(a,b) to figure \ref{fig:rankd_error}(a,b). By increasing the basis size to $d=10$, the final truncation errors are noticeably reduced, but adding more vectors does not lead to further reduction of the already low errors. For the spillway case shown in figure \ref{fig:rankd_error_mode5}(c,d), the effect of noise in the data becomes apparent once more. Here, increasing the basis size from $d=10$ to $d=25$ reduces both the eigenvalue and eigenvector errors. At the same time, however, the earlier comparison of the mode shapes in figure \ref{fig:spillway_compare_modes} demonstrated that the coherent large-scale structures are accurately captured, even for $d=5$. 

The truncation error analysis suggests that the basis size $d$ should be chosen somewhat larger than the desired number of modes. It is also important to emphasize that the definitions of the truncation errors rely on the batch solution as a reference, which itself may not be statistically fully converged.

\section{Discussion}\label{discussion}

In this work, we introduce an algorithm that recursively updates the SPOD of large or streaming datasets. In \S\S \ref{errors}-\ref{rankderror}, we demonstrate that the algorithm is capable of converging the most energetic SPOD modes while lifting the requirement to store potentially prohibitively large amounts of data.

The batch algorithm requires storage of $n_{x}\times\nvar\times n_t$
data points plus another $n_{x}\times\nvar\times\left(\frac{\nfreq}{2}+1\right)\times\nblk$ points for the spectral estimation of a real signal. In its simplest implementation, all data is loaded into memory simultaneously. If the data set is too large to be stored in memory, the $\nblk$ temporal Fourier transforms can be computed a priori and stored, fully or partly, on hard drive, and then be reloaded and processed frequency by frequency. The drawbacks of this approach are the significantly longer computing time resulting from the read/write operations, and the additional storage requirements. For higher-dimensional data, e.g.~three-dimensional fields, snapshots totaling multiple terabytes are likely to be required to converge the second-order statistics, in particular those of subdominant modes. In such cases, batch SPOD may become computationally intractable altogether. 

The streaming SPOD algorithm, on the other hand, has a much lower storage requirement of $n_{x}\times\nvar\times n_{\mathrm{freq}}^{\prime} + n_{x}\times\nvar\times \left(\frac{n_{\mathrm{freq}}^{\prime}}{2}+1 \right) \times d$ data points for $\Xmat_k^{(m)}$ and $\Umat_k^{(m)}$, respectively, plus a number of small fields that do not scale with the large dimensions in space and time. Here, $n_{\mathrm{freq}}^{\prime} \leq \nfreq$ is the number of frequencies to be analyzed. Ideally, $\Xmat_k^{(m)}$ and $\Umat_k^{(m)}$ are stored and updated in memory during runtime. Besides its low storage requirements, the algorithm achieves computational efficiency by employing MGS steps for the orthogonalization.

A useful implication of the ergodicity assumption is that it offsets the need to store a single long time-series. In \S \ref{onlinespod}, we used overlapping blocks to increase the number of Fourier samples in cases where the total number of snapshots is limited. A quite different scenario occurs when dealing with fast data streams. In such a scenario, we can take advantage of the fact that ergodicity permits arbitrarily long gaps between sampling periods of two consecutive data blocks $\Xmat_k^{(m)}$ and $\Xmat_k^{(m+1)}$. This is advantageous in experimental setups such as time-resolved particle image velocimetry (TR-PIV). It suffices to utilize $\nfreq$ consecutive snapshots at a time, perform the computationally costly cross-correlations to obtain velocity data, and update the SPOD eigenbasis before continuing to sample data. This procedure allows, in principle, to converge second-order flow statics over arbitrarily long time horizons.

\paragraph{Acknowledgements}
Support of the Office of Naval Research grant No. N00014-16-1-2445 with Dr. Knox Millsaps as program manager is gratefully acknowledged. Special thanks are due to Patrick Vogler for reviewing the manuscript and sharing his insights, and to Tim Colonius, Andres Goza and Matthias Kramer for making valuable comments. The author gratefully acknowledges Matthias Kramer and Hubert Chanson for providing the experimental optical flow data. The experiments were undertaken in the hydraulics laboratory at the University of Queensland. The LES study was supported by NAVAIR SBIR project, under the supervision of Dr. John T. Spyropoulos. The main LES calculations were carried out on CRAY XE6 machines at DoD HPC facilities in ERDC DSRC.

\section*{References}
\bibliographystyle{elsarticle-num}
\bibliography{jets,acoustics,others,non-modalStab,mypublications_proceedings,mypublications_journals,PODetc}

\begin{thebibliography}{10}
\expandafter\ifx\csname url\endcsname\relax
  \def\url#1{\texttt{#1}}\fi
\expandafter\ifx\csname urlprefix\endcsname\relax\def\urlprefix{URL }\fi
\expandafter\ifx\csname href\endcsname\relax
  \def\href#1#2{#2} \def\path#1{#1}\fi

\bibitem{taira2017modal}
K.~Taira, S.~L. Brunton, S.~Dawson, C.~W. Rowley, T.~Colonius, B.~J. McKeon,
  O.~T. Schmidt, S.~Gordeyev, V.~Theofilis, L.~S. Ukeiley, AIAA Journal. \href
  {http://dx.doi.org/https://doi.org/10.2514/1.J056060}
  {\path{doi:https://doi.org/10.2514/1.J056060}}.

\bibitem{rowley2017model}
C.~W. Rowley, S.~T. Dawson,
  Annual Review of Fluid Mechanics 49 (2017) 387--417.

\bibitem{sirovich1987turbulence}
L.~Sirovich, Quarterly of
  applied mathematics 45~(3) (1987) 561--571.

\bibitem{aubry1988dynamics}
N.~Aubry, P.~Holmes, J.~L. Lumley, E.~Stone, Journal of Fluid
  Mechanics 192 (1988) 115--173.

\bibitem{noack2003hierarchy}
B.~R. Noack, K.~Afanasiev, M.~Morzy{\'n}ski, G.~Tadmor, F.~Thiele, Journal of Fluid Mechanics 497 (2003) 335--363.

\bibitem{Lumley:1970}
J.~L. Lumley, New York, 1970.

\bibitem{towneschmidtcolonius_2018_jfm}
A.~Towne, O.~T. Schmidt, T.~Colonius, Journal of Fluid Mechanics 847 (2018) 821–867.
\newblock \href {http://dx.doi.org/10.1017/jfm.2018.283}
  {\path{doi:10.1017/jfm.2018.283}}.

\bibitem{schmid2010dmd}
P.~J. {Schmid}, Journal of Fluid Mechanics 656 (2010) 5--28.
\newblock \href {http://dx.doi.org/10.1017/S0022112010001217}
  {\path{doi:10.1017/S0022112010001217}}.

\bibitem{hemati2014dynamic}
M.~S. Hemati, M.~O. Williams, C.~W. Rowley, Physics of Fluids 26~(11) (2014) 111701.

\bibitem{zhang2017online}
H.~Zhang, C.~W. Rowley, E.~A. Deem, L.~N. Cattafesta, arXiv preprint arXiv:1707.02876.

\bibitem{businger1970updating}
P.~Businger,BIT 10~(3) (1970)
  376--385.

\bibitem{de1985svd}
R.~D. De~Groat, R.~A. Roberts, in: Circuits, Systems and Computers, 1985. Nineteeth Asilomar
  Conference on, IEEE, 1985, pp. 601--605.

\bibitem{brand2006fast}
M.~Brand, Linear algebra and its applications 415~(1) (2006) 20--30.

\bibitem{brand2002incremental}
M.~Brand, Proc. Eur. Conf. Computer Vision (2002) 707--720.

\bibitem{brand2003fast}
M.~Brand, in:
  Proceedings of the 2003 SIAM International Conference on Data Mining, SIAM,
  2003, pp. 37--46.

\bibitem{turney2010frequency}
P.~D. Turney, P.~Pantel, Journal of artificial intelligence research 37 (2010) 141--188.

\bibitem{braconnier2011towards}
T.~Braconnier, M.~Ferrier, J.-C. Jouhaud, M.~Montagnac, P.~Sagaut, Computers \& Fluids
  40~(1) (2011) 195--209.

\bibitem{Solomon1991}
O.~M. {Solomon}, Jr., NASA STI/Recon
  Technical Report N 92.

\bibitem{bresetal_2018jfm}
G.~A. Br{\`e}s, P.~Jordan, V.~Jaunet, M.~Le~Rallic, A.~V.~G. Cavalieri,
  A.~Towne, S.~K. Lele, T.~Colonius, O.~T. Schmidt, Journal of Fluid
  Mechanics 851 (2018) 83--124.
\newblock \href {http://dx.doi.org/10.1017/jfm.2018.476}
  {\path{doi:10.1017/jfm.2018.476}}.

\bibitem{Kramer2017}
M.~Kramer, H.~Chanson, Environmental Fluid
  Mechanics (2018) 1--19.

\bibitem{Zhang2017}
G.~Zhang, H.~Chanson, Experimental Thermal and Fluid Science 90~(July 2017) (2017)
  186--199.

\bibitem{glauser1987coherent}
M.~N. Glauser, S.~J. Leib, W.~K. George, Turbulent shear flows 5 (1987)
  134--145.

\bibitem{SchmidtEtAl_2017_JFM}
O.~T. Schmidt, A.~Towne, T.~Colonius, A.~V.~G. Cavalieri, P.~Jordan, G.~A.
  Brès, Journal of Fluid Mechanics 825
  (2017) 1153–1181.
\newblock \href {http://dx.doi.org/10.1017/jfm.2017.407}
  {\path{doi:10.1017/jfm.2017.407}}.

\bibitem{bres2017unstructured}
G.~A. Br{\`e}s, F.~E. Ham, J.~W. Nichols, S.~K. Lele, AIAA Journal 55~(4) (2017) 1164--1184.

\bibitem{SchmidtEtAl_2018_JFM}
O.~T. Schmidt, A.~Towne, G.~Rigas, T.~Colonius, G.~A. Br{\`e}s, Journal of Fluid Mechanics 855 (2018) 953–982.
\newblock \href {http://dx.doi.org/10.1017/jfm.2018.675}
  {\path{doi:10.1017/jfm.2018.675}}.

\bibitem{Liu2015}
T.~Liu, A.~Merat, M.~H.~M. Makhmalbaf, C.~Fajardo, P.~Merati, Experiments in Fluids 56~(8) (2015) 1--23.

\end{thebibliography}

\end{document}